\documentclass[10pt,a4paper]{article}
\pdfoutput=1
\usepackage{jheppub}
\usepackage{feynmp-auto}
\usepackage[utf8]{inputenc}
\usepackage[fleqn]{mathtools}
\usepackage[fleqn]{amsmath}
\usepackage{amssymb,mathrsfs}
\usepackage{setspace}
\usepackage{graphicx,floatflt,rotate, cancel}
\usepackage{bm}
\usepackage{color}
\usepackage{xcolor}
\usepackage{graphicx}
\usepackage{multirow}
\usepackage{array}
\usepackage{float}
\usepackage[toc,page]{appendix}
\usepackage{diagbox}
\usepackage[mathscr]{euscript}
\usepackage{cleveref}
\usepackage{soul}
\usepackage{pifont}
\usepackage{subfig}
\usepackage[numbers]{natbib}
\usepackage{notoccite}

\makeatletter
\gdef\@fpheader{}
\makeatother

\usepackage{tikz}
\usepackage{scalerel}

\definecolor{lime}{HTML}{A6CE39}
\DeclareRobustCommand{\orcidicon}{\hspace{-2.1mm}
\begin{tikzpicture}
\draw[lime,fill=lime] (0,0.0) circle [radius=0.13] node[white] {{\fontfamily{qag}\selectfont \tiny ID}}; \draw[white,fill=white] (-0.0525,0.095) circle [radius=0.007]; 
\end{tikzpicture} \hspace{-3.7mm} }

\newcommand{\orcidC}[1]{\href{https://orcid.org/#1}{\orcidicon}}

\newcommand{\orcidauthorC}{0000-0001-6587-951X}

\definecolor{orange}{rgb}{1,0.5,0}

\title{Probing the Inert Doublet Dark Matter with Stellar-Mass Black Hole Mini-Spikes}

\author{Rameswar Sahu\orcidC{\orcidauthorC}}
\emailAdd{rameswarsahu1@gmail.com}
\affiliation{Department of Physics and Astrophysics, University of Delhi, Delhi 110007, India}
\affiliation{Department of Physics, SGTB Khalsa College, Delhi 110007, India}


\abstract{
The nature of dark matter remains a central unresolved problem in contemporary physics, motivating the exploration of well-defined extensions of the Standard Model. Among these, the Inert Doublet Model provides a minimal and theoretically consistent framework accommodating a viable weakly interacting massive particle dark matter candidate. In this work, we investigate the IDM parameter space through an analysis of \textit{Fermi}-LAT observations of dark matter mini-spikes surrounding stellar-mass black holes. Owing to the strong gravitational compression of dark matter in the vicinity of these systems, the resulting annihilation signal can be significantly enhanced, rendering such environments exceptionally sensitive probes of dark matter interactions. We find that substantial regions of the IDM parameter space, particularly in the high-mass regime, are subject to stringent constraints extending into the multi-TeV range. These results underscore the increasingly important role of indirect detection in probing particle dark matter scenarios beyond the reach of current collider and direct detection experiments.
}

\begin{document}

\maketitle
\flushbottom


\section{Introduction}
\label{sec:intro}

The existence of dark matter (DM) is firmly established through a wide range of astrophysical and cosmological observations, including galactic rotation curves, gravitational lensing, large-scale structure formation, and precision measurements of the cosmic microwave background (CMB) \cite{Rubin:1970zza, Fixsen:1996nj, Planck:2018vyg, Primack:1997av, Clowe:2003tk}. Despite this overwhelming gravitational evidence, the microscopic nature of DM remains unknown (see e.g.~\cite{Bertone:2004pz,Arcadi:2017kky,Arcadi:2024ukq, Schumann:2019eaa} for some recent reviews). In particular, the Standard Model (SM) of particle physics does not contain a viable particle capable of accounting for the observed relic abundance, $\Omega_{\rm DM} h^2 \simeq 0.12$, as measured by the \textit{Planck} collaboration~\cite{Planck:2018vyg}\footnote{Standard Model neutrinos constitute a hot (or at best warm) dark matter component and cannot simultaneously account for the observed relic abundance and structure formation \cite{Tremaine:1979we, Bond:1980ha}.}. This deficiency provides strong motivation for extensions of the Standard Model that naturally accommodate stable and cosmologically viable dark matter candidates.

Among the simplest and most extensively studied realizations of weakly interacting massive particle (WIMP) dark matter is the Inert Doublet Model (IDM). The model extends the SM scalar sector with an additional $\mathrm{SU}(2)_L$ doublet charged under a discrete $\mathbb{Z}_2$ symmetry, under which all SM fields remain even. As a consequence, the lightest neutral component of the inert doublet becomes stable and can serve as a viable dark matter candidate. Since its inception in Ref.~\cite{Deshpande:1977rw} and its subsequent recognition as a viable dark matter framework in Ref.~\cite{Ma:2006km,Barbieri:2006dq}, the model has been subjected to extensive scrutiny in light of constraints from the measured relic abundance~\cite{LopezHonorez:2006gr, LopezHonorez:2010tb, Goudelis:2013uca, Dolle:2009fn,Tytgat:2007cv, Arina:2009um,Eiteneuer:2017hoh}, direct and indirect DM searches~\cite{Goudelis:2013uca, LopezHonorez:2010tb, Arina:2009um, Gustafsson:2007pc, Garcia-Cely:2015khw, Eiteneuer:2017hoh, Justino:2024etz,Klasen:2013btp}, and collider experiments~\cite{Belyaev:2018ext, Lundstrom:2008ai, Dolle:2009ft, Ilnicka:2015jba,Miao:2010rg, Datta:2016nfz, Belyaev:2016lok, Braathen:2024ckk}. Although these probes place significant restrictions on the model parameter space, substantial viable regions still survive.

A persistent challenge in the phenomenological study of dark matter within BSM frameworks lies in our incomplete understanding of the thermal history of the Universe prior to Big Bang nucleosynthesis (BBN). Conventional relic density computations assume a standard cosmological paradigm, characterized by instantaneous reheating, conservation of entropy in the visible sector, and a radiation-dominated expansion governed by the Standard Model energy density. However, well-motivated departures from this picture can arise naturally, giving rise to non-standard cosmological histories~\cite{Allahverdi:2020bys, Batell:2024dsi}. In such scenarios, the predicted dark matter relic abundance can deviate substantially from its standard counterpart. For instance, an extended reheating phase can lead to a significant dilution of the relic density if DM freeze-out occurs during reheating \cite{Belanger:2024yoj, Gelmini:2006pw, Bernal:2022wck, Bernal:2024yhu,Mondal:2025awq}, whereas an epoch dominated by a stiff fluid with equation-of-state parameter $w > 1/3$ enhances the Hubble expansion rate, thereby increasing the freeze-out abundance relative to the standard case \cite{Salati:2002md, Profumo:2003hq} (see, e.g., Ref.~\cite{Belanger:2025ack} for a recent analysis in the context of the IDM). These considerations introduce an intrinsic cosmological uncertainty in mapping relic density measurements onto constraints in the model parameter space. This motivates the study of complementary probes that are largely insensitive to assumptions regarding the pre-BBN cosmological evolution.

Indirect detection provides one such avenue. Since the annihilation rate of WIMP dark matter scales with the square of the local dark matter density, environments with highly concentrated dark matter distributions can generate observable annihilation signals \cite{Slatyer:2021qgc, Bertone:2004pz, Jungman:1995df}. In this context, stellar-mass black holes (sBHs) have recently emerged as particularly promising targets \cite{Ireland:2024lye}. The adiabatic growth \cite{Gondolo:1999ef,Ireland:2024lye} of a black hole can gravitationally compress the surrounding dark matter distribution, leading to the formation of dense mini-spikes \cite{Gondolo:1999ef, Sadeghian:2013laa} that substantially enhance the expected gamma-ray flux. Recent analyses based on 17 years of observations by the \textit{Fermi} Large Area Telescope (LAT) \cite{Fermi-LAT:2009ihh} have exploited this mechanism to derive remarkably stringent limits on the dark matter annihilation cross section from the absence of gamma-ray excesses around stellar-mass black holes~\cite{Braga:2026frt}.

Motivated by these developments, in the present work we investigate the implications of such indirect detection constraints for the IDM. Building upon the framework developed in Ref.~\cite{Braga:2026frt}, we recast the \textit{Fermi}-LAT limits from dark matter mini-spikes surrounding stellar-mass black holes into constraints on the IDM parameter space. Importantly, we do not require the IDM dark matter candidate to reproduce the observed relic abundance, allowing us to assess the indirect detection constraints independently of the standard thermal freeze-out assumption. We find that the resulting bounds probe the IDM well into the multi-TeV regime, excluding dark matter masses up to $\mathcal{O}(15)$~TeV depending on the model parameters. These constraints significantly extend beyond the reach of current direct detection searches and far exceed the kinematic capabilities of present and near-future collider experiments, demonstrating the exceptional potential of compact astrophysical systems in probing heavy dark matter scenarios.

The remainder of this article is organized as follows. In Sec.~\ref{sec:model}, we briefly review the Inert Doublet Model. Sec.~\ref{sec:constraints} summarizes the relevant theoretical and experimental constraints on the model parameter space. In Sec.~\ref{sec:dm}, we discuss the dark matter phenomenology of the model and present the constraints arising from direct and indirect detection searches, including the primary results of this work. Finally, we summarize our findings and conclude in Sec.~\ref{sec:con}.


\section{The Inert Doublet Model}
\label{sec:model}

The IDM is one of the simplest extensions of the Standard Model (SM) capable of accommodating a viable WIMP dark matter candidate. The model enlarges the scalar sector of the SM through the introduction of an additional $\mathrm{SU}(2)_L$ scalar doublet, $\Phi$, and imposes a discrete $\mathbb{Z}_2$ symmetry under which $\Phi$ is odd while all SM fields are even. As a consequence, the inert doublet does not couple directly to SM fermions through Yukawa interactions. Moreover, if the $\mathbb{Z}_2$ symmetry remains unbroken after electroweak symmetry breaking, the lightest $\mathbb{Z}_2$-odd state is stable and can therefore constitute a cosmologically viable DM candidate.

The scalar sector of the model is described by the $\mathbb{Z}_2$-symmetric potential
\begin{equation}
\begin{aligned}
    V_0 &= \mu_1^2 |H|^2 + \mu_2^2 |\Phi|^2 
    + \lambda_1 |H|^4 + \lambda_2 |\Phi|^4 
    + \lambda_3 |H|^2 |\Phi|^2 
    + \lambda_4 |H^\dagger \Phi|^2 \\
    &\quad + \frac{\lambda_5}{2} \left[ (H^\dagger \Phi)^2 + \mathrm{H.c.} \right],
\end{aligned}
\label{Eq:TreePotential}
\end{equation}
where $H$ denotes the SM Higgs doublet and $\Phi$ the inert scalar doublet. Without loss of generality, the Lagrangian parameters can be taken to be real, such that the scalar potential conserves CP.

After electroweak symmetry breaking (EWSB), the scalar fields can be parametrized as
\begin{equation}
    H = \frac{1}{\sqrt{2}}
    \begin{pmatrix}
        \sqrt{2}\, G^+ \\
        v + h + i G^0
    \end{pmatrix},
    \qquad
    \Phi = \frac{1}{\sqrt{2}}
    \begin{pmatrix}
        \sqrt{2}\, H^+ \\
        H^0 + i A^0
    \end{pmatrix},
\end{equation}
where $G^\pm$ and $G^0$ denote the would-be Goldstone bosons, $h$ is the SM-like Higgs boson with vacuum expectation value $v \simeq 246~\mathrm{GeV}$, and $H^\pm$, $H^0$, and $A^0$ are the physical inert scalars. In the CP-conserving limit considered here, $H^0$ and $A^0$ correspond to CP-even and CP-odd neutral states, respectively. The $\mathbb{Z}_2$ symmetry is preserved if only the Higgs doublet acquires a vacuum expectation value and $\langle \Phi \rangle = 0$. This corresponds to the so-called inert vacuum, which is phenomenologically required to ensure DM stability.

At tree level, the scalar mass spectrum is given by
\begin{align}
    m_h^2 &= 2 \lambda_1 v^2, \\
    m_{H^0}^2 &= \mu_2^2 + \lambda_L v^2, \label{Eq:mH0tree} \\
    m_{A^0}^2 &= \mu_2^2 + \lambda_S v^2, \\
    m_{H^\pm}^2 &= \mu_2^2 + \frac{1}{2} \lambda_3 v^2,
\end{align}
where we have defined
\begin{equation}
    \lambda_L \equiv \frac{1}{2}(\lambda_3 + \lambda_4 + \lambda_5),
    \qquad
    \lambda_S \equiv \frac{1}{2}(\lambda_3 + \lambda_4 - \lambda_5).
\end{equation}

The minimization condition of the potential fixes $\mu_1^2 = -\lambda_1 v^2$, while $\lambda_1$ is determined by the observed Higgs mass. It is convenient to trade the remaining Lagrangian parameters for physical masses and couplings. In particular, $\mu_2^2$, $\lambda_3$, $\lambda_4$, and $\lambda_5$ can be expressed as
\begin{align}
\mu_2^2 &= m_{H^0}^2 - \lambda_L v^2, \\
\lambda_3 &= \frac{2}{v^2} \left( m_{H^\pm}^2 - m_{H^0}^2 + \lambda_L v^2 \right), \\
\lambda_4 &= \frac{m_{A^0}^2 + m_{H^0}^2 - 2 m_{H^\pm}^2}{v^2}, \\
\lambda_5 &= \frac{m_{H^0}^2 - m_{A^0}^2}{v^2}.
\end{align}

A convenient set of independent parameters is therefore given by
\begin{equation}
    \{ m_{H^0},\, m_{A^0},\, m_{H^\pm},\, \lambda_L,\, \lambda_2 \}.
\end{equation}
Among these, $\lambda_L$ controls the Higgs-portal interaction and thus plays a central role in DM phenomenology, while $\lambda_2$ affects the model only indirectly through theoretical constraints such as perturbativity, vacuum stability, and unitarity. In this work, we fix $\lambda_2 = 0.1$ as a representative choice.

Depending on the mass hierarchy, either $H^0$ or $A^0$ can serve as the DM candidate. Since their phenomenology is largely analogous, we focus on the $H^0$ DM scenario without loss of generality. In the following, we will discuss the relevant theoretical and experimental constraints on the IDM parameter space.


\section{Constraints}
\label{sec:constraints}

The parameter space of the IDM is subject to a variety of theoretical and experimental constraints, which have been extensively studied in the literature~\cite{Ilnicka:2015jba,Cao:2007rm,Agrawal:2008xz,Gustafsson:2007pc,Dolle:2009fn,Dolle:2009ft,LopezHonorez:2006gr,Arina:2009um,Tytgat:2007cv,LopezHonorez:2010eeh,Krawczyk:2009fb,Kanemura:1993hm,Akeroyd:2000wc,Swiezewska:2012ej,Lundstrom:2008ai,Gustafsson:2010zz,Gustafsson:2012aj,Modak:2015uda}. In this work, we impose the subset of constraints most relevant for the dark matter phenomenology considered here. Unless stated otherwise, the theoretical consistency conditions and electroweak precision observables are evaluated using {\tt 2HDMC}~\cite{Eriksson:2009ws}.

\subsection{Theoretical constraints}

We first require the scalar potential to satisfy the standard theoretical consistency conditions. In particular, the potential must remain bounded from below in all field directions, thereby ensuring the absence of runaway instabilities at large field values. Perturbative unitarity is imposed by requiring the tree-level scalar $2\to2$ scattering amplitudes to satisfy the usual partial-wave unitarity bounds. In addition, all quartic couplings are required to remain within the perturbative regime.

A further condition arises from the requirement that the electroweak vacuum corresponds to the inert minimum, such that the $\mathbb{Z}_2$ symmetry remains unbroken after electroweak symmetry breaking. This is necessary to guarantee the stability of the dark matter candidate. Following Refs.~\cite{Ginzburg:2010wa,Gustafsson:2010zz,Swiezewska:2012ej}, we impose the condition
\begin{equation}
\label{eq:invac}
\frac{m_{11}^2}{\sqrt{\lambda_1}}
\geq
\frac{m_{22}^2}{\sqrt{\lambda_2}},
\end{equation}
which ensures that the inert vacuum constitutes the global minimum of the scalar potential.

\subsection{Experimental constraints}

We next impose the relevant collider and precision constraints on the model parameter space. The SM-like Higgs boson is required to reproduce the observed mass,
\begin{equation}
    m_h \simeq 125~\mathrm{GeV},
\end{equation}
in agreement with LHC measurements~\cite{ATLAS:2015yey}. We additionally require the total Higgs width to satisfy $\Gamma_h < 22~\mathrm{MeV}$~\cite{CMS:2014quz,ATLAS:2015cuo}, which constrains invisible and exotic Higgs decay modes involving inert scalars whenever such channels are kinematically accessible.

Electroweak precision observables further place stringent restrictions on the inert scalar spectrum through their contributions to the oblique parameters $(S,T,U)$~\cite{Altarelli:1990zd,Peskin:1990zt,Peskin:1991sw,Maksymyk:1993zm}. Following Ref.~\cite{Ilnicka:2015jba}, we impose the condition
\begin{equation}
\chi^2_{\mathrm{STU}}
=
\mathbf{x}^T
\mathbf{C}^{-1}
\mathbf{x}
\leq
8.025,
\end{equation}
where
\begin{equation}
\mathbf{x}^T
=
(S-\hat S,\,
T-\hat T,\,
U-\hat U),
\end{equation}
with hatted quantities denoting the experimentally measured central values, the unhatted quantities representing the model predictions  and $\mathbf{C}$ denote the corresponding covariance matrix. The numerical inputs are taken from Refs.~\cite{Baak:2014ora,Ilnicka:2015jba}.

Collider limits on additional scalar states are implemented using {\tt HiggsBounds-5.10.2}~\cite{Bechtle:2008jh,Bechtle:2011sb,Bechtle:2013wla}, while compatibility with Higgs signal strength measurements is enforced through {\tt HiggsSignals-2.6.2}~\cite{Bechtle:2013xfa}.

Additional constraints arise from precision measurements of electroweak gauge boson widths~\cite{ParticleDataGroup:2024cfk} and from reinterpretations of LEP searches for inert scalars~\cite{Lundstrom:2008ai,EspiritoSanto:2003by}. These bounds primarily affect the low-mass region of the model with inert scalar masses below $\sim100~\mathrm{GeV}$. All the above constraints have been consistently taken into account in our analysis.


\section{Dark Matter Phenomenology}
\label{sec:dm}

In the Inert Doublet Model, the lightest neutral $\mathbb{Z}_2$-odd scalar provides a viable weakly interacting massive particle (WIMP) dark matter candidate. In this work, we identify $H^0$ as the dark matter particle. The relic abundance of $H^0$ is determined by the interplay between particle physics processes governing its interactions with the Standard Model and the other inert scalars, and the cosmological expansion of the Universe. These effects enter through the Boltzmann equation,
\begin{equation}
    \frac{dn}{dt} + 3 H n = - \langle \sigma v \rangle \left(n^2 - n_{\mathrm{eq}}^2 \right),
\end{equation}
where $n$ is the number density of $H^0$, $n_{\mathrm{eq}}$ its equilibrium value, $H$ the Hubble expansion rate, and $\langle \sigma v \rangle$ the thermally averaged annihilation cross section.

From the particle physics perspective, the dark matter phenomenology of the IDM is controlled by a small set of parameters: the dark matter mass $m_{H^0}$, the Higgs-portal coupling $\lambda_L$, and the mass splitting between $H^0$ and the heavier inert scalars. In the following, we consider a degenerate spectrum for the heavier states, $m_{A^0} = m_{H^\pm}$, and define $\Delta M \equiv m_{A^0} - m_{H^0}$. The Higgs portal interaction, proportional to $\lambda_L$, governs annihilation into fermions and Higgs bosons, while gauge interactions determine annihilation into electroweak gauge bosons. In addition, coannihilation processes involving $A^0$ and $H^\pm$ become important when the mass splitting $\Delta M$ is small, significantly modifying the effective annihilation rate.

The dark matter phenomenology of the IDM has been extensively studied in the literature under the assumption of a standard radiation-dominated cosmological history \cite{LopezHonorez:2006gr, LopezHonorez:2010tb, Goudelis:2013uca, Dolle:2009fn,Tytgat:2007cv, Arina:2009um,Eiteneuer:2017hoh}. In this framework, the observed relic abundance, $\Omega_{\mathrm{DM}} h^2 \simeq 0.12$, can be understood in terms of distinct mass regimes characterized by different dominant annihilation channels. For $m_{H^0} \lesssim m_W$, annihilation proceeds predominantly through the Higgs portal into fermion pairs, with a rate controlled by $\lambda_L$. In this region, a resonant enhancement occurs when $m_{H^0} \simeq m_h/2$, corresponding to the Higgs funnel, where the observed relic density can be achieved even for relatively small values of the portal coupling.

As the dark matter mass approaches the electroweak scale, annihilation into off-shell and subsequently on-shell gauge bosons becomes increasingly important. Once the $W^+W^-$ channel opens, gauge interactions dominate and typically lead to a large annihilation cross section, resulting in an underabundant relic density over most of this mass range. Nevertheless, interference effects between different contributions can partially suppress the total annihilation rate, allowing for viable parameter space in restricted regions \cite{LopezHonorez:2010tb}. 

For $m_{H^0} \gtrsim m_h$, annihilation into Higgs boson pairs becomes kinematically accessible and further enhances the annihilation efficiency, reinforcing the tendency toward underabundance \cite{Goudelis:2013uca}. However, at higher masses, $m_{H^0} \gtrsim 500~\mathrm{GeV}$, the observed relic density can again be reproduced \cite{Belanger:2025ack}. In this regime, the interplay between different annihilation channels reduces the effective annihilation cross section and lead to a viable high-mass region.

The above picture relies on the assumption of a standard thermal history prior to Big Bang Nucleosynthesis. However, the expansion history of the Universe at earlier times is not directly constrained, allowing for well-motivated deviations from radiation domination \cite{Allahverdi:2020bys, Batell:2024dsi}. In scenarios with a prolonged reheating phase, for instance, entropy injection from the decay of a heavy field can dilute the dark matter abundance \cite{Belanger:2024yoj, Gelmini:2006pw, Bernal:2022wck, Bernal:2024yhu,Mondal:2025awq}, leading to relic densities that are significantly smaller than those predicted in the standard cosmology. Conversely, if the same field couples to the dark sector, late-time production of dark matter can enhance the relic abundance after freeze-out \cite{Gelmini:2006pq, Cline:2026cek}. Similarly, cosmological histories with a faster expansion rate, such as kination-like scenarios, can increase the relic density by reducing the efficiency of annihilation processes \cite{Salati:2002md, Profumo:2003hq}. In fact, a recent study~\cite{Belanger:2025ack} has demonstrated that, in kination-like non-standard cosmological scenarios, the intermediate-mass region of the IDM (120 GeV $< M_{H^0} <$ 550 GeV )—typically underabundant in the standard cosmology—can be rendered consistent with the observed dark matter relic density.

These considerations imply that the relic density prediction is inherently sensitive to assumptions about the pre-BBN cosmological history. As a consequence, the requirement of reproducing the observed dark matter abundance cannot be regarded as an absolute constraint on the parameter space of the model in a model-independent manner. In this work, we therefore adopt a conservative approach and do not impose the relic density constraint as a strict exclusion criterion. Instead, we consider the full parameter space consistent with the theoretical and experimental constraints discussed in Sec.~\ref{sec:constraints}, and subsequently examine the implications of direct and indirect detection searches.

Finally, we specify the region of parameter space that will be the focus of our analysis. In the low-mass regime, $m_{H^0} \lesssim 100~\mathrm{GeV}$, a large fraction of the parameter space is already tightly constrained by direct and indirect detection searches \cite{Eiteneuer:2017hoh}, with only narrow viable regions remaining, notably near the Higgs resonance and around $m_{H^0} \sim 70~\mathrm{GeV}$. Moreover, recent studies~\cite{Belanger:2025ack} have shown that the intermediate-mass range, $100~\mathrm{GeV} \lesssim m_{H^0} \lesssim 225~\mathrm{GeV}$, is strongly constrained by gamma-ray observations of dwarf spheroidal galaxies, particularly from \textit{Fermi}-LAT data. Nevertheless, for completeness, our indirect detection analysis is performed over the full mass range $m_{H^0} > 60~\mathrm{GeV}$, including regions already subject to strong existing constraints.

The IDM model files used in our analysis were implemented using \texttt{FeynRules}~\cite{Alloul:2013bka}. All numerical calculations presented in this work, including the evaluation of the spin-independent dark matter--nucleon scattering cross section and the dark matter annihilation spectra relevant for indirect detection, have been performed using \texttt{micrOMEGAs~6.0}~\cite{Alguero:2023zol}.


\subsection{Direct Detection}

Before turning to the primary focus of this work, namely the indirect detection constraints derived from \textit{Fermi}-LAT observations of stellar-mass black holes, we briefly discuss the current status of direct detection bounds on the IDM parameter space. As will become evident in the following subsection, indirect detection probes the high-mass regime of the model far beyond the reach currently attainable in direct detection experiments. For this reason, our discussion here is intended primarily to provide a qualitative overview of the direct detection phenomenology of the IDM. Consequently, although current direct detection experiments retain sensitivity to dark matter masses extending up to $\mathcal{O}(10~\mathrm{TeV})$, we restrict the presentation of bounds to the range $100~\mathrm{GeV} \lesssim m_{H^0} \lesssim 1~\mathrm{TeV}$, since, as we demonstrate in the next subsection, the entire region $100~\mathrm{GeV} \lesssim m_{H^0} \lesssim 10~\mathrm{TeV}$ is excluded by indirect detection constraints. Therefore, extending the presentation of direct detection limits beyond the range shown here would not modify the phenomenological conclusions of this work.

In the IDM, spin-independent dark matter--nucleon scattering arises predominantly through Higgs exchange at tree level, with the interaction strength controlled by the Higgs-portal coupling $\lambda_L$. The corresponding scattering cross section scales as
\begin{equation}
    \sigma_{\mathrm{SI}} \propto \lambda_L^2,
\end{equation}
and is therefore insensitive to the sign of $\lambda_L$. Gauge interactions do not contribute to elastic scattering at tree level. In particular, the absence of a diagonal $ZH^0H^0$ coupling forbids potentially large $Z$-mediated elastic scattering processes, which would otherwise lead to severe experimental constraints.

The model nevertheless allows inelastic scattering processes mediated by the $Z$ boson, such as
\begin{equation}
    H^0 N \rightarrow A^0 N.
\end{equation}
However, these processes become kinematically inaccessible once a sufficiently large mass splitting is introduced between the inert neutral scalars. Throughout this work, we impose a minimum splitting
\begin{equation}
    \Delta M \equiv m_{A^0}-m_{H^0} \gtrsim 1~\mathrm{GeV}.
\end{equation}
This choice guarantees the prompt decay of the charged scalar $H^\pm$, thereby avoiding constraints from long-lived particle searches at the LHC, including disappearing track and heavy stable charged particle searches \cite{Heisig:2018kfq, Alguero:2021dig}. This choice also makes the contribution from the inelastic scattering processes subdominant.

At the loop level, electroweak corrections induce additional contributions to the scattering amplitude, leading to a non-vanishing lower bound on the direct detection cross section even in the limit $\lambda_L \to 0$. These effects have been already studied in the literature \cite{Klasen:2013btp}. In the present work, however, all dark matter observables are computed consistently at tree level, and loop-induced contributions are therefore not included in our analysis.

\begin{figure}
    \centering
    \includegraphics[width=0.5\linewidth]{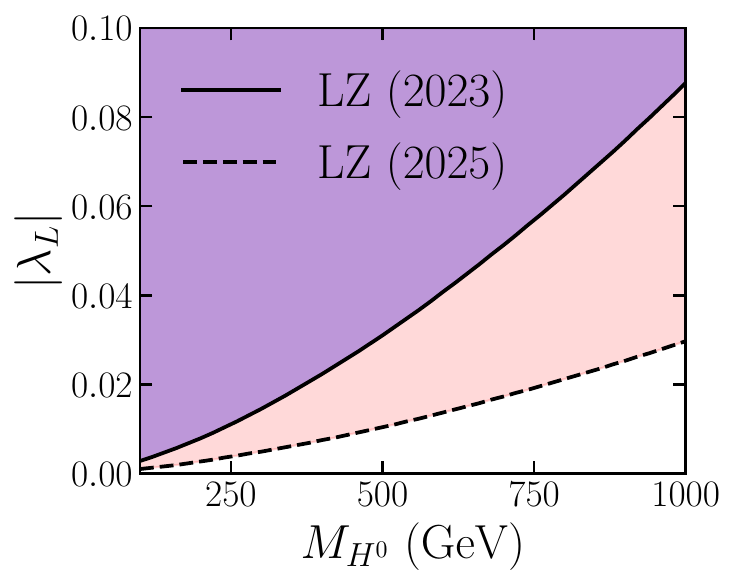}
    \caption{Maximum value of $\lambda_L$ allowed by DM direct detection.}
    \label{fig:dm_DD}
\end{figure}

To assess the impact of present direct detection experiments, we employ the latest limits from the LUX-ZEPLIN (LZ) collaboration~\cite{LZ:2024zvo}. The resulting constraints are shown in Fig.~\ref{fig:dm_DD}. As a validation of our implementation, we first reproduce the exclusion region obtained using the 2023 LZ dataset, shown in purple, and find excellent agreement with the results reported in Ref.~\cite{Belanger:2025ack}.

The LZ results place stringent constraints on the Higgs-portal interaction across the sub-TeV region of the IDM parameter space. In particular, the 2023 data excludes $|\lambda_L| \gtrsim 0.1$, for DM mass near $1~\mathrm{TeV}$. The updated 2025 results further strengthen these bounds, reaching sensitivities of $|\lambda_L| \gtrsim \mathcal{O}(0.02)$ for dark matter masses near the TeV scale, as indicated by the pink exclusion region in Fig.~\ref{fig:dm_DD}. 

Having briefly summarized the implications of present direct detection searches, we now turn to the central result of this work: the constraints on the IDM parameter space derived from \textit{Fermi}-LAT observations of stellar-mass black hole mini-spikes.

\subsection{Indirect Detection}
\label{subsec:indirect}

Indirect detection offers a complementary probe of dark matter by searching for the stable products of its annihilation in astrophysical environments. The strength of such signals depends sensitively on the dark matter density, as the annihilation rate scales with its square, making compact and highly concentrated systems particularly attractive targets \cite{Slatyer:2021qgc}. In this context, stellar-mass black holes (sBHs) can act as efficient enhancers of annihilation signals. The gravitational potential of a black hole can significantly distort the surrounding dark matter distribution, increasing the density in its vicinity and thereby enhancing the expected gamma-ray flux. 

In this work, we adopt the framework developed in Ref.~\cite{Braga:2026frt}, in which the dark matter distribution around stellar-mass black holes is modeled under the assumption of adiabatic growth \cite{Gondolo:1999ef,Ireland:2024lye}. In this scenario, the slow evolution of the gravitational potential leads to a compression of the ambient dark matter, resulting in a steep density profile in the vicinity of the black hole \cite{Gondolo:1999ef, Sadeghian:2013laa}. At sufficiently small radii, this growth is limited by dark matter annihilation, which imposes a maximal (saturation) density~\cite{Sadeghian:2013laa}. At larger radii, tidal interactions with the Galactic environment can truncate the outer regions of the spike~\cite{Chanda:2022hls}. Despite these effects, the annihilation signal is dominated by the innermost region, where the density is highest and relatively insensitive to the details of the outer profile~\cite{Braga:2026frt}.

The gamma-ray flux from dark matter annihilation can be factorized into a particle physics component, determined by the annihilation cross section and final states, and an astrophysical contribution encoded in the $J$-factor. In this work, we adopt the analytical expressions for the $J$-factor derived in Ref.~\cite{Braga:2026frt}, which relate it to the properties of the black hole and the underlying dark matter parameters. The details of this derivation are presented in Ref.~\cite{Braga:2026frt} and are not repeated here.

To confront the predicted gamma-ray flux with observational data, we follow the analysis strategy of Ref.~\cite{Braga:2026frt}, which is based on 17 years of observations from the \textit{Fermi} Large Area Telescope (LAT)~\cite{Fermi-LAT:2009ihh}. The analysis focuses on the black hole low-mass X-ray binaries A0620--00 and XTE~J1118+480, which are treated as point-like sources given the compactness of the expected emission region. A binned likelihood analysis is performed using the \texttt{fermipy} framework~\cite{Wood:2017yyb}, built upon the \texttt{Fermitools} suite, with standard event selection and background modeling. The analysis is carried out over the energy range from 500~MeV to 1~TeV, including contributions from Galactic diffuse emission, an isotropic background, and all cataloged sources within the region of interest. In the absence of a significant excess, the resulting likelihood profiles for the gamma-ray flux are used to place upper limits on the annihilation signal. We refer the reader to Ref.~\cite{Braga:2026frt} for further details of the analysis.

To validate our implementation, we follow the setup of Ref.~\cite{Braga:2026frt} and compute the theoretical differential gamma-ray flux using the dark matter annihilation spectra from the PPPC4DMID framework~\cite{Cirelli:2010xx} for three representative annihilation channels, $b\bar{b}$, $\tau^+\tau^-$, and $W^+W^-$. Note that, Ref.~\cite{Braga:2026frt} finds comparable constraints on the annihilation cross section for the two stellar-mass black holes considered. Therefore, in the following, we only focus on the source XTE~J1118+480, without loss of generality. 

We reproduce the benchmark results presented in Table~III of Ref.~\cite{Braga:2026frt}, finding excellent agreement across the parameter space. This provides a non-trivial validation of both the astrophysical modeling and the statistical analysis pipeline employed in this work. The comparison is summarized in Table~\ref{tab:dm_limits}, where the results from Ref.~\cite{Braga:2026frt} are shown alongside our reproduction.

\begin{table}[htbp]
\centering
\caption{95\% C.L. upper limits on the dark matter annihilation cross section, $\langle \sigma v \rangle$ (in cm$^3\,\mathrm{s}^{-1}$), for the stellar-mass black hole XTE~J1118+480. Values shown in parentheses correspond to our reproduced results, while those outside parentheses are taken from Ref.~\cite{Braga:2026frt}.}
\label{tab:dm_limits}
\resizebox{\columnwidth}{!}{%
\begin{tabular}{l|c|c|c}
\hline
\textbf{Channel} & \textbf{100 GeV} & \textbf{1 TeV} & \textbf{10 TeV} \\
\hline 

\hline
$W^+W^-$       & $8.73(8.69) \times 10^{-32}$ & $8.46(6.51) \times 10^{-31}$ & $9.64(7.76) \times 10^{-29}$ \\
$b\bar{b}$     & $2.63(3.31) \times 10^{-32}$ & $3.11(2.41) \times 10^{-31}$ & $2.88(2.27) \times 10^{-29}$ \\
$\tau^+\tau^-$ & $6.36(4.71) \times 10^{-32}$ & $1.75(1.67) \times 10^{-29}$ & $1.75 (1.59) \times 10^{-25}$ \\
\hline
\end{tabular}%
}
\end{table}

We then apply this framework to the IDM. The photon spectra are computed using \texttt{micrOMEGAs~6.0}~\cite{Alguero:2023zol}, which provides the differential yields $\mathrm{d}N/\mathrm{d}E$ including contributions from all kinematically accessible annihilation channels. These spectra are combined with the $J$-factors described above to obtain the predicted gamma-ray flux for each point in the model parameter space.

To derive constraints, we compare the predicted flux with the \textit{Fermi}-LAT likelihood profiles obtained above. For a fixed dark matter mass, we scan over the annihilation cross section $\langle\sigma v\rangle$ and compute the corresponding likelihood by summing the contributions from all energy bins. Following the standard approach in gamma-ray analyses, we define the test statistic as $TS = -2\,\Delta\ln\mathcal{L}$, where $\Delta\ln\mathcal{L}$ is the difference between the log-likelihood at a given value of $\langle\sigma v\rangle$ and its maximum value. In the absence of a significant excess, upper limits on $\langle\sigma v\rangle$ are obtained by requiring $ TS = 2.71$, which corresponds to a one-sided 95\% confidence level for one degree of freedom. 

\begin{figure}
    \centering
    \includegraphics[width=0.45\linewidth]{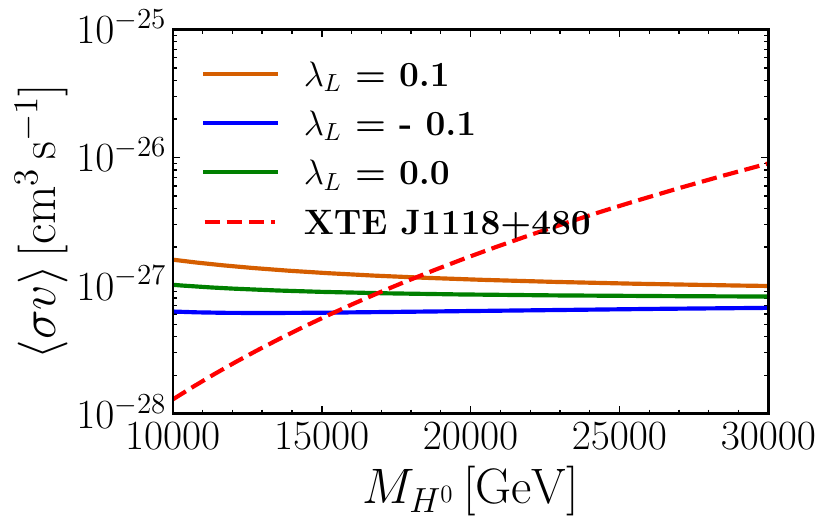}
    \includegraphics[width=0.45\linewidth]{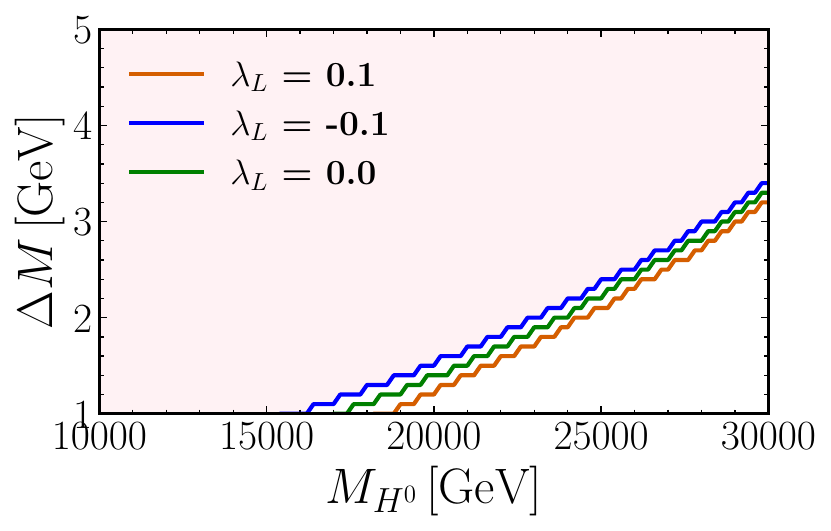}
    \caption{ Left: 95\% C.L. upper limits on the dark matter annihilation cross section from \textit{Fermi}-LAT observations (red), compared with the IDM predictions for $\lambda_L = 0.1$, $0.0$, and $-0.1$. Right: Constraints on the IDM parameter space in the $M_{H^0}$--$\Delta M$ plane for the same benchmark values of $\lambda_L$.}
    \label{fig:dm_IDD}
\end{figure}

The resulting constraints on the annihilation cross section are presented in Fig.~\ref{fig:dm_IDD} and constitute the central result of this work. Note that since the bounds on $\lambda_L$ mostly come from DD searches and the currently DD searches are largely insensitive for DM masses above 10 TeV we have only shown results for three qualitative values of $\lambda_L$. The left panel displays the 95\% C.L. upper limits on $\langle\sigma v\rangle$ derived from \textit{Fermi}-LAT observations of the stellar-mass black hole XTE~J1118+480 (red dashed line), together with the corresponding theoretical predictions of the IDM. The latter are shown for a fixed mass splitting $\Delta M = 1~\mathrm{GeV}$ and for three representative values of the Higgs-portal coupling, $\lambda_L = -0.1$, $0.0$, and $0.1$, depicted by the blue, green, and orange solid curves, respectively. As evident from the figure, dark matter masses of approximately $15.5~\mathrm{TeV}$, $17~\mathrm{TeV}$, and $18~\mathrm{TeV}$ are excluded in these three scenarios. 

The right panel illustrates the dependence of the upper limits on the mass splitting $\Delta M$ for the same benchmark choices of $\lambda_L$. One observes that, over a wide mass range extending up to $\mathcal{O}(30~\mathrm{TeV})$, only highly degenerate spectra remain viable. Although Fig.~\ref{fig:dm_IDD} displays results only in the mass range $10~\mathrm{TeV} < M_{H^0} < 30~\mathrm{TeV}$ for clarity, our analysis was performed over the full interval $60~\mathrm{GeV} < M_{H^0} < 30~\mathrm{TeV}$. We find that the entire low- and intermediate-mass region, $60~\mathrm{GeV} < M_{H^0} < 10~\mathrm{TeV}$, is excluded by the \textit{Fermi}-LAT constraints for all values of $\Delta M$ and $\lambda_L$. Taken together, these results highlight the remarkable reach of indirect detection in probing the IDM parameter space. In particular, the constraints derived here extend well beyond the sensitivity of current direct detection experiments, which lose sensitivity for dark matter masses above $\sim 10~\mathrm{TeV}$, and far exceed the kinematic reach of present and near-future collider searches. Consequently, indirect detection emerges as a uniquely powerful probe of the high-mass regime of the IDM, playing a dominant role in testing this class of models.


\section{Conclusion}
\label{sec:con}

The Inert Doublet Model (IDM) constitutes one of the most economical and theoretically consistent extensions of the Standard Model capable of accommodating a weakly interacting massive particle dark matter candidate. In this work, we have examined the viability of its parameter space in light of recent indirect detection constraints derived from gamma-ray observations of stellar-mass black holes, with particular emphasis on the system XTE~J1118+480. These constraints, inferred from 17 years of \textit{Fermi}-LAT data, exploit the presence of dark matter mini-spikes formed via gravitational compression, which substantially enhance the annihilation signal and render such systems exceptionally sensitive probes of particle dark matter.

Building upon the analysis framework of Ref.~\cite{Braga:2026frt}, we implemented the corresponding likelihood formalism and validated our pipeline through an explicit reproduction of benchmark results. We then recast these constraints within the IDM by computing the annihilation-induced gamma-ray flux across the model parameter space and confronting it with the observational likelihood profiles.

Our results demonstrate that indirect detection bounds from stellar-mass black holes impose stringent constraints on the IDM, excluding dark matter masses up to approximately $15.5~\mathrm{TeV}$, $17~\mathrm{TeV}$, and $18~\mathrm{TeV}$ for representative values of the Higgs-portal coupling $\lambda_L = -0.1$, $0.0$, and $0.1$, respectively, under the assumption of a quasi-degenerate inert scalar spectrum with $\Delta M = 1~\mathrm{GeV}$. Furthermore, allowing for variations in the mass splitting reveals that, over a wide mass range extending to $\mathcal{O}(30~\mathrm{TeV})$, only highly degenerate configurations remain consistent with observational limits. 

These findings underscore the remarkable reach of indirect detection in probing the high-mass regime of the IDM, significantly extending beyond the sensitivity of current direct detection experiments and far surpassing the kinematic capabilities of present and near-future collider searches. In this regime, astrophysical observations emerge not merely as complementary tools, but as indispensable probes of particle dark matter.

More broadly, our analysis illustrates the power of leveraging compact astrophysical environments to test particle physics models in regions of parameter space that are otherwise difficult to access. We anticipate that similar strategies may yield comparably strong constraints across a wide class of beyond-the-Standard-Model scenarios featuring annihilating dark matter candidates, thereby reinforcing the synergistic interplay between astrophysics and particle physics in the ongoing effort to unravel the nature of dark matter.

\acknowledgments The work of RS is supported by ANRF CRG/2023/008234.

\bibliography{v0}
\bibliographystyle{JHEP}
\end{document}